\documentclass[12pt,a4paper,landscape]{revtex4}
\usepackage{epsfig}                           
\begin{document}
\title{Classical  and Thermodynamic work fluctuations}
\author{Mamata Sahoo}
\author{A.M Jayannavar}
\address{Institute of Physics,~~Sachivalaya Marg,~~ Bhubaneswar-751005, India}
\begin{abstract}
Abstract:~~We have studied the nature of classical work ($W_{c}$) and thermodynamic work ($W$) fluctuations in  systems driven out of equilibrium both in transient and time periodic steady state. As the observation time of trajectory increases, we show that the number of trajectories which exhibit excursions away from the typical behaviour i.e., $W_{c}<0$, $W<\Delta F$ and dissipated heat $Q<0$ decreases as anticipated for macroscopic time scales. Analytical expressions for such trajectories are obtained. Trajectory for which $W_{c}<0$ may not correspond to $W<\Delta F$ or $Q<0$. The applicability of steady state fluctuation theorems are discussed in our linear as well as nonlinear models. 
\end{abstract}
\maketitle
\maketitle

\vspace{3.5cm}
\noindent
\newcommand{\nwc}{\newcommand}
\nwc{\bdm}{\begin{displaymath}}
\nwc{\edm}{\end{displaymath}}
\nwc{\pd}[2]{\frac{\partial #1}{\partial #2}}

Corresponding Author:  A.M. Jayannavar\\
Email address       :  jayan@iopb.res.in
\newpage
\section{Introduction}
Over last decade, nonequilibrium fluctuation theorems have attracted much interest. They provide relations for physical quantities such as work, heat and entropy in driven nonequilibrium systems, independent of the nature of driving [1-17]. The fluctuation relations are statements about the symmetry of the distributions of the physical quantities around zero and not around the maximum. These distribution functions exhibit finite weight for negative values for physical quantities of interest, which are usually rare and are related to transient second law violating contributions at microscopic length scales. Fluctuation theorems quantify the probability of those non-equilibrium trajectories, taken individually, violate some of the inequalities of the thermodynamics. For average quantities, these theorems lead to inequalities consistent with the second law of thermodynamics. In our present study, we are mainly concerned with Jarzynski Equality (JE) [10,11] for thermodynamic work ($W$) and Bochkov-Kuzovlev identity (BKI) for classical work ($W_{c}$)[14-17]. JE relates the nonequilibrium work done with equilibrium free energy difference ($\Delta F=F_{B}-F_{A}$) between two thermodynamic states. These two thermodynamic states are uniquely characterized by the initial (A) and final (B) values of the time dependent protocol. JE is given by 
\begin{equation}
\langle e^ {-\beta W} \rangle=e^{-\beta \Delta F} ,
\label{Jar}
\end{equation}
 where $W$ is the thermodynamic work. The angular bracket $\langle \cdots  \rangle$ denotes the average over an ensemble of realizations, starting from initial equilibrium  configurations. $W$ is defined as $W=\int_{0}^{\tau} \frac{\partial U(x,t)}{\partial t} dt$,  where $U(x,t)$ is the effective potential of the system. Incidentally, the JE finds its counterpart in the relation known as Bochkov-Kuzovlev identity which has been proposed much earlier [14-17]
\begin{equation}  
 \langle e^{-\beta W_{c}} \rangle=1
 \label{Buz}
 \end{equation}
  where  $W_{c}$ is the classical work. If the potential is decomposed as $U(x,t)=U_{0}(x)+U_{p}(x,t)$, the classical work done over a time interval $\tau$ is defined as $W_{c}=-\int_{0}^{\tau} \frac{\partial U_{p}(x,t)}{\partial x} \dot{x} dt$.~~~For the validity of BKI,~~the initial equilibrium distribution must correspond to the time independent potential ($P_{e}(x_{0})=N                                                                                                                                                                                                                                                                                                e^{-\frac{U_{0}(x_{0})}{k_{B}T}}$, $N$ being the normalization constant). Seifert has shown that both JE and BKI are special cases of more general result, obtained within a framework that defines entropy production in nonequilibrium state [18,19]. When the internal energy of the system is defined by the bare Hamiltonian/potential, then classical work is defined as the work performed by the application of an external force that affects the system in a fixed landscape of the bare Hamiltonian. To define thermodynamic work, external time dependent perturbation is treated as the time dependent contribution to the internal energy of the system [20]. The work done by the system on the external bodies which produce a change in the Hamiltonian is known as thermodynamic work $W$. It turns out that $W$ is more useful than $W_{c}$ for the reconstruction of free energy landscapes. Using Jensen's inequality, from eqns. (1) and (2) it follows that $\langle W \rangle \geq \Delta F$, $\langle W_{c} \rangle \geq 0$, which are consistent with the physics at macroscopic scales. Observations of realizations wherein $W < \Delta F$ or $W_{c}< 0$ are treated as excursions away from the typical macroscopic behaviour. The finite time trajectories for which $W < \Delta F$, are sometimes referred to as transient second law violating trajectories [21]. In our work, we have also studied decrease in the number of such trajectories as a function of time of observations and some pertinent questions are raised. Nature of the fluctuations in $W$ and $W_{c}$ are  studied in a nonlinear system exhibiting stochastic resonance(SR) [22-29]. Finally we have analyzed the validity of SSFT for $W$ and $W_{c}$.

\section{Driven linear systems}
\flushleft{
\subsection{Model}}
\normalsize

We consider a linear model, i.e.,~ a overdamped Brownian particle in a one dimensional harmonic potential $U_{0}(x)=\frac{1}{2} k x^{2}$ in the presence of an external time dependent potential $U_{p}=-A x(t)  \sin(\omega t+\phi)$, where $A$ , $\omega$ are the amplitude, frequency of the external drive respectively and  $\phi$ is an initial phase. The dynamics of the particle is described by the Langevin's equation [30] 
\begin{equation}
\gamma \frac{dx}{dt}=-\frac{\partial U(x,t)}{\partial x}+\xi(t) ,
\label{Lang}
\end{equation}
where $U(x,t)=U_{0}(x)+U_{p}(x,t)$.~~The random force field $\xi(t)$ is a zero mean Gaussian white noise, i.e., $\langle \xi(t)\xi(t^{'}) \rangle=2 D \delta(t-t^{'})$, $D=\gamma k_{B}T$ is the noise strength of the medium. Here $\gamma$ is the friction coefficient, $k_{B}$ is the Boltzmann constant and $T$ is the absolute temperature of the bath. In the following we use a dimensionless form of eqn.(3),~~namely 
 
 \begin{equation}
\frac{dx}{dt}=-\frac{\partial U(x,t)}{\partial x}+\xi(t) ,
\label{Lang1}
\end{equation}
All the calculated quantities are in dimensionless form.
\subsection{Thermodynamic work distributions:}
Thermodynamic work($W$) done on the system by the external drive over a  time $t$ is given 

\begin{equation}
W=-A \omega\int_{0}^{t}  x(t^{'}) \cos(\omega t^{'}+\phi) dt^{'}.
\label{therm}
\end{equation}
 The formal solution of eqn.(4) is
\begin{equation}
x(t)=x_{0}\exp(-kt) +\int_{0}^{t} dt^{'} \exp(-k(t-t^{'})) [A\sin(\omega t^{'}+\phi)+\xi(t^{'})].
\label{ini}
\end{equation}

Where $x_{0}$ is the initial coordinate of the particle. The initial distribution for $x_{0}$ is assumed to be the equilibrium canonical distribution, $P_{e}(x_{0})=\sqrt{\frac{k\beta }{2 \pi}} \exp(-\frac{k\beta}{2}(x_{0}-\frac{A}{k}\sin(\phi))^{2})$. From eqns.(5) and (6), it follows that  thermodynamic work done is a  linear functional of the Gaussian variable $\xi(t)$ . Hence the distribution $P(W)$ of work $W$  is a  Gaussian [31-33]. By a simple algebra one can evaluate $P(W)$ analytically. The full  probability distribution $P(W)$ is given by
\begin{eqnarray}
P(W)=\frac{1}{\sqrt{2\pi \sigma^2}} \exp{\left[-\frac{(W-\langle W \rangle)^{2}}{2\sigma^{2}}\right]}
\end{eqnarray}

where the analytical expressions for the mean thermodynamic work done($\langle W \rangle$) is given in Appendix-A.
The variance  $\sigma^{2}=\langle W^{2} \rangle -\langle W \rangle^{2}$  is given by 
\begin{eqnarray}
\langle W^2 \rangle - \langle W \rangle^2 &=& \frac{2}{\beta} \langle W \rangle +\frac{A^{2}}{k \beta}\sin^{2}(\omega t+\phi).
\end{eqnarray}
The thermodynamic work done over a time interval $\tau$ satisfies JE,
\begin{eqnarray}
\langle e^{-\beta W} \rangle =e^{-\beta \Delta F}
\end{eqnarray}
where $\Delta F=-\frac{A^2}{2k}\sin^{2}(\omega \tau+\phi)$,~~is the free energy difference between two thermodynamic states with potentials $U(x_{0})=\frac{1}{2} kx_{0}^{2}-Ax_{0}\sin(\phi)$ and $U(x,\tau)=\frac{1}{2} kx^{2}-Ax(\tau)\sin(\omega \tau+\phi)$ respectively .

We furthur study the statistics of the thermodynamic work done, $W$, in the
time asymptotic regime (i.e, in the limit $t \rightarrow \infty$). In this limit, probability distributions are time periodic with a period of magnitude $\frac{2\pi}{\omega}$.~~The average work done($\langle W_{\tau} \rangle$) over a time of observation $\tau$ in the time periodic steady state is
\begin{eqnarray}
\langle W_{\tau} \rangle 
&=& \lim_{ t\rightarrow \infty}[\langle W(t+\tau) \rangle-\langle W(t) \rangle]\nonumber\\ &=&\frac{A^{2} \omega^{2} \tau}{2(k^2+\omega^2)} .  
\end{eqnarray}
The variance($\sigma_{\tau}^{2}$) of $W_{\tau}$ averaged over a time of observation $\tau$ is given by 
\begin{eqnarray}
\sigma_{\tau}^{2}&=&\langle W_{\tau}^2 \rangle-\langle W_{\tau} \rangle^2\nonumber\\
&=&\frac{2}{\beta}\langle W_{\tau} \rangle +\Delta(\tau) 
\end{eqnarray}
The analytical expression for $\Delta(\tau)$ is given in Appendix-C. The distribution for $W_{\tau}$ is Gaussian. However, it does not satisfy the fluctuation dissipation relation, namely $\sigma_{\tau}^{2}=\frac{2}{\beta}\langle W_{\tau} \rangle$ [4]. Hence SSFT is not valid over small time of observation. However, $\langle W_{\tau} \rangle$ is an extensive quantity in $\tau$, whereas $\Delta(\tau)$ saturates to a finite value for $\tau \gg \frac{1}{k}$. For large $\tau$, $\frac{2}{\beta}\langle W_{\tau} \rangle\gg \Delta(\tau)$ and the contribution of $\Delta(\tau)$ to the variance $\sigma_{\tau}^{2}$ can be ignored in eqn.(11). Under this approximation SSFT holds. Here SSFT implies   

\begin{eqnarray}
\frac{P(W_{\tau})}{P(-W_{\tau})}=e^{\beta W_{\tau}}
\end{eqnarray}

\subsection{Classical work distributions:}
Classical work ($W_{c}$) done on the system by the external drive $A\sin(\omega t+\phi)$ over a duration of time $t$ is given by [14,15]
\begin{eqnarray}
W_{c}&=&\int_{0}^{t} A\sin(\omega t^{'}+\phi) \dot{x}(t^{'}) dt^{'} \nonumber\\
&=& W+\left[A\sin(\omega t+\phi)x(t)-A\sin(\phi)x(0)\right] 
\label{cl}
\end{eqnarray}
Note that classical work ($W_{c}$) differs from the thermodynamic work($W$) by a boundary term.~~From eqns.(13) and (6) it follows that $W_{c}$ is a linear functional of  a Gaussian variable $\xi(t)$,~~consequently $P(W_{c})$ is a Gaussian and is given by
 \begin{eqnarray}
P(W_{c})=\frac{1}{\sqrt{2\pi \sigma_{c}^2}} \exp\left[-\frac{(W_{c}-\langle W_{c} \rangle)^{2}}{2\sigma_{c}^{2}}\right]
\end{eqnarray}
where $\langle W_{c} \rangle$ is the mean classical work  done over a time interval $t$ and  $\sigma_{c}^{2}$ is the fluctuation or variance of $W_{c}$.~~~These quantities can be evaluated analytically. 
The expression for average or mean classical work($\langle W_{c} \rangle$) is given in Appendix-B.

The variance in classical work ($\sigma_{c}^{2}$) over a transient time $t$ is given by
\begin{eqnarray}
\sigma_{c}^{2}&=&\langle W_{c}^2 \rangle - \langle W_{c} \rangle^2 \nonumber\\
 &=& \frac{2}{\beta} \langle W_{c} \rangle
\end{eqnarray}
To calculate $W_{c}$ and $\sigma_{c}^{2}$,~~we assumed initially the system to be in equilibrium with distribution $P_{e}(x_{0})=\sqrt(\frac{k\beta}{2\pi}) e^{-\beta \frac{kx_{0}^{2}}{2}}$. The expression for $P(W_{c})$ is consistent with BKI.
\begin{equation}
\langle e^{-\beta W_{c}} \rangle=1.
\end{equation} 
The average classical work ($\langle W_{c\tau} \rangle$) done over a time of observation $\tau$ in the time periodic steady state, is given by
\begin{eqnarray}
\langle W_{c\tau} \rangle =\frac{A^{2}\omega^{2}\tau}{2(k^2+\omega^2)},
\end{eqnarray}
which has the same magnitude  as that of $\langle W_{\tau} \rangle$ and is independent of $\phi$. The distribution $P(W_{c\tau})$ is Gaussian. The variance ($\sigma_{c\tau)}$) is given by 
\begin{eqnarray}
\sigma_{c\tau}=\frac{2}{\beta}\langle W_{c\tau} \rangle+\Delta_{1}(\tau)
\end{eqnarray}
An expression for $\Delta_{1}(\tau)$ is given in Appendix-D. For $\tau \gg \frac{1}{k}$, $\Delta_{1}(\tau)$ saturates to a finite value. For large $\tau$, for which $\frac{2}{\beta} \langle W_{c\tau} \rangle \gg \Delta_{1}(\tau)$ one can ignore $\Delta_{1}(\tau)$ in eqn.(18). Under this condition the classical work satisfies SSFT.
\vskip 1cm
\begin{figure}[htp!]
\begin{center}
\input{epsf}
\includegraphics [width=4.5in,height=3.5in] {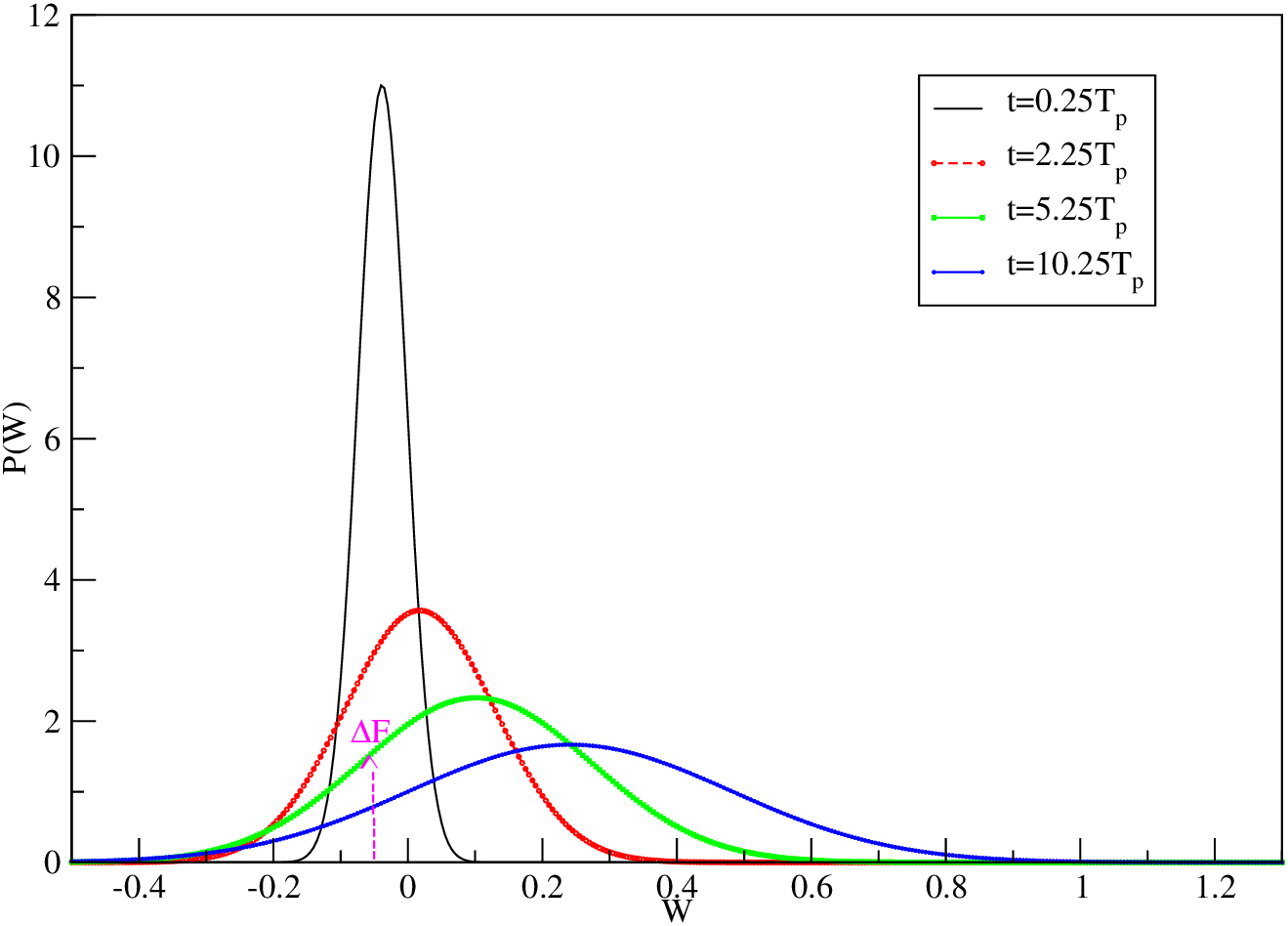}
\caption{The distribution $P(W)$ as a function of transient time $t$ (in units of $T_{p}$) for parameters $A=0.3$, $D=0.1$, $\omega=0.1$ and $T_{p}=\frac{2\pi}{\omega}$.} 
\end{center}
\end{figure}

In fig(1) and (2), we have plotted the probability distributions of $W$ and $W_{c}$ for various values of time periods in the transient regime using our analytical results (eqns.(7),(14)). Other parameters are given in figure captions. For short time $t=0.25T_{p}$ ($T_{p}$ being the time period of magnitude $\frac{2\pi}{\omega}$), we notice that most of the weight of $W$ is located in the negative side. The weight towards negative side for values $W<\Delta F$ comes from the so called transient second law violating trajectories [21]. As we increase the time of observations, the total weight for $W<\Delta F$ (i.e., area under the curve $P(W)$ between $W=-\infty$ and $W=\Delta F$) decreases (see fig(1)). We would like to emphasize that depending on the protocol and potential, one may obtain peak in $P(W)$ at values $W<\Delta F$, i.e., most probable value of $W$ is inconsistent with the classical macroscopic or thermodynamic results. However, $\langle W \rangle >\Delta F$ is always satisfied. In contrast, $W_{c}$ does not exhibit a peak in the negative region. We have checked this separately for different systems for different protocols.
\begin{figure}[htp!]
\begin{center}
\input{epsf}
\includegraphics [width=4.5in,height=3.5in] {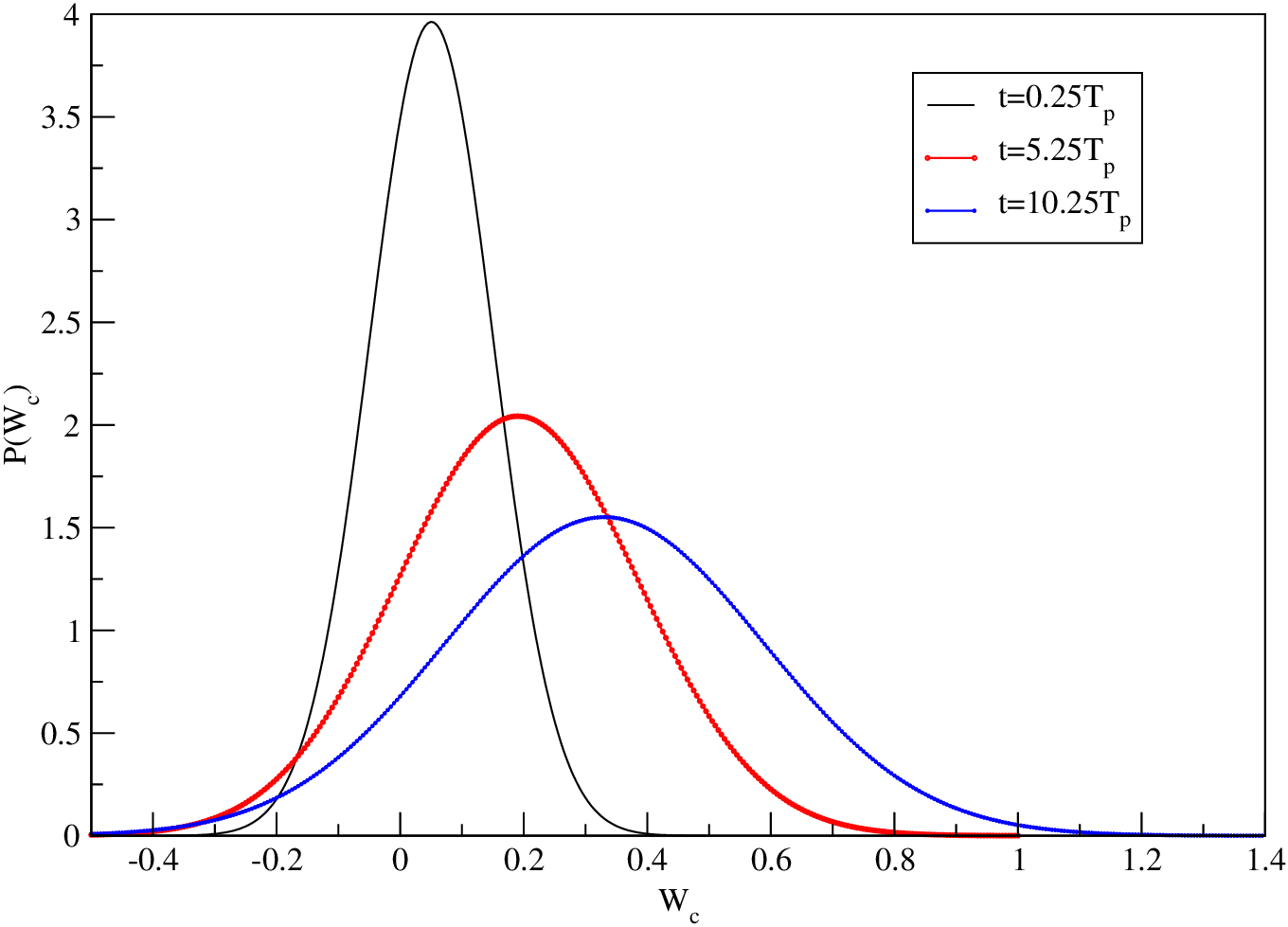}
\caption{The distribution $P(W_{c})$ as a function of transient time $t$ (in units of $T_{p}$) for parameters $A=0.3$, $D=0.1$, $\omega=0.1$ and $T_{p}=\frac{2\pi}{\omega}$.} 
\end{center}
\end{figure}

The physical quantities such as work ($W$), heat ($Q$), total entropy ($\Delta S_{tot}$) and internal energy ($\Delta U$) can be calculated using the method of stochastic energetics [34]. For details, we refer to [26,27,28,35]. 
\begin{figure}[htp!]
\begin{center}
\input{epsf}
\includegraphics [width=4.5in,height=3.5in] {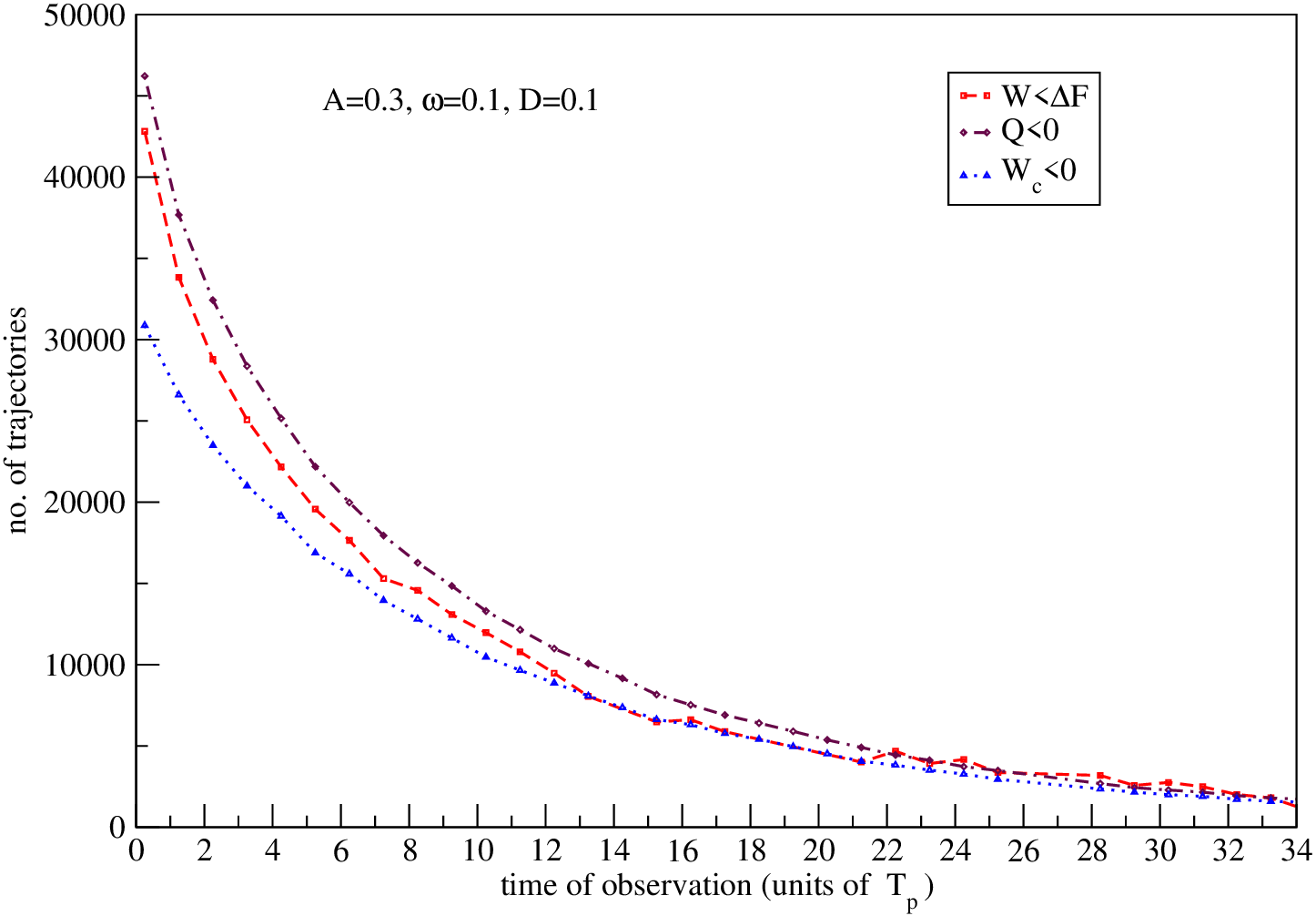}
\caption{ Number of trajectories for $W<0$, $Q<0$ and $W_{cl}<0$ as a function of time $t$ (in units of $T_{p}$). The physical parameters are same as in fig(1) and (2).} 
\end{center}
\end{figure}
In fig(3) we have plotted number of trajectories which do not satisfy our conventional wisdom at macroscopic scale, namely $W<\Delta F$, $W_{c}<0$ and heat dissipated, $Q<0$. For each individual trajectory, $W$, $W_{c}$ and $Q$ are calculated. In our simulation, we have generated $10^{5}$ trajectories. Number of trajectories for which $W_{c}<0$, $W<\Delta F$ and $Q<0$ are all different. Trajectory for which $W_{c}<0$ need not correspond to $W<\Delta F$ or $Q<0$. It can be readily shown that if $N$ is the total number of observed trajectories for $W_{c}$, then the number of trajectories $N^{'}$ for which $W_{c}<0$ are given by $N^{'}=\frac{N}{2} \textrm{Erfc}(\frac{\langle W_{c} \rangle}{\sqrt{2\sigma_{W_{c}}^{2}}})$. In the time asymptotic regime $\langle W_{c} \rangle$ scales with $t$ and the decay of $N^{'}$ with time $t$ is given by $N^{'}=\frac{a}{\sqrt{t}}\exp(-ct)$, where $a$ and $c$ are constants. For our present problem $a=\sqrt{\frac{2D(k^{2}+\gamma^{2}\omega^{2})}{A^{2} \gamma \omega^{2} \pi}}$ and $c=\frac{A^{2} \gamma \omega^{2}}{8D(k^{2}+\gamma^{2}\omega^{2})}$. The number of trajectories for which $W<\Delta F$ in the large time limit, decay with the same functional form as that for $W_{c}<0$. Hence there is no correlation amongst trajectories in regard to the transient violations in $W$, $W_{c}$ and $Q$. For our simple linear problem and protocol (up to time periods $\approx 12$), number of trajectories for which $Q<0$ are greater than $W<\Delta F$ which in turn are greater than $W_{c}<0$. This is not a general rule. Depending on the system and protocol, different possibilities exist. However, one common observation is that the number of trajectories which defy our general classical notion for $W$, $W_{c}$ and $Q$ decrease monotonically as time of observation is increased, leading to classical thermodynamic behaviour in the time asymptotic regime as is evident from fig(3).
\begin{figure}[htp!]
\begin{center}
\input{epsf}
\includegraphics [width=4.5in,height=3.5in] {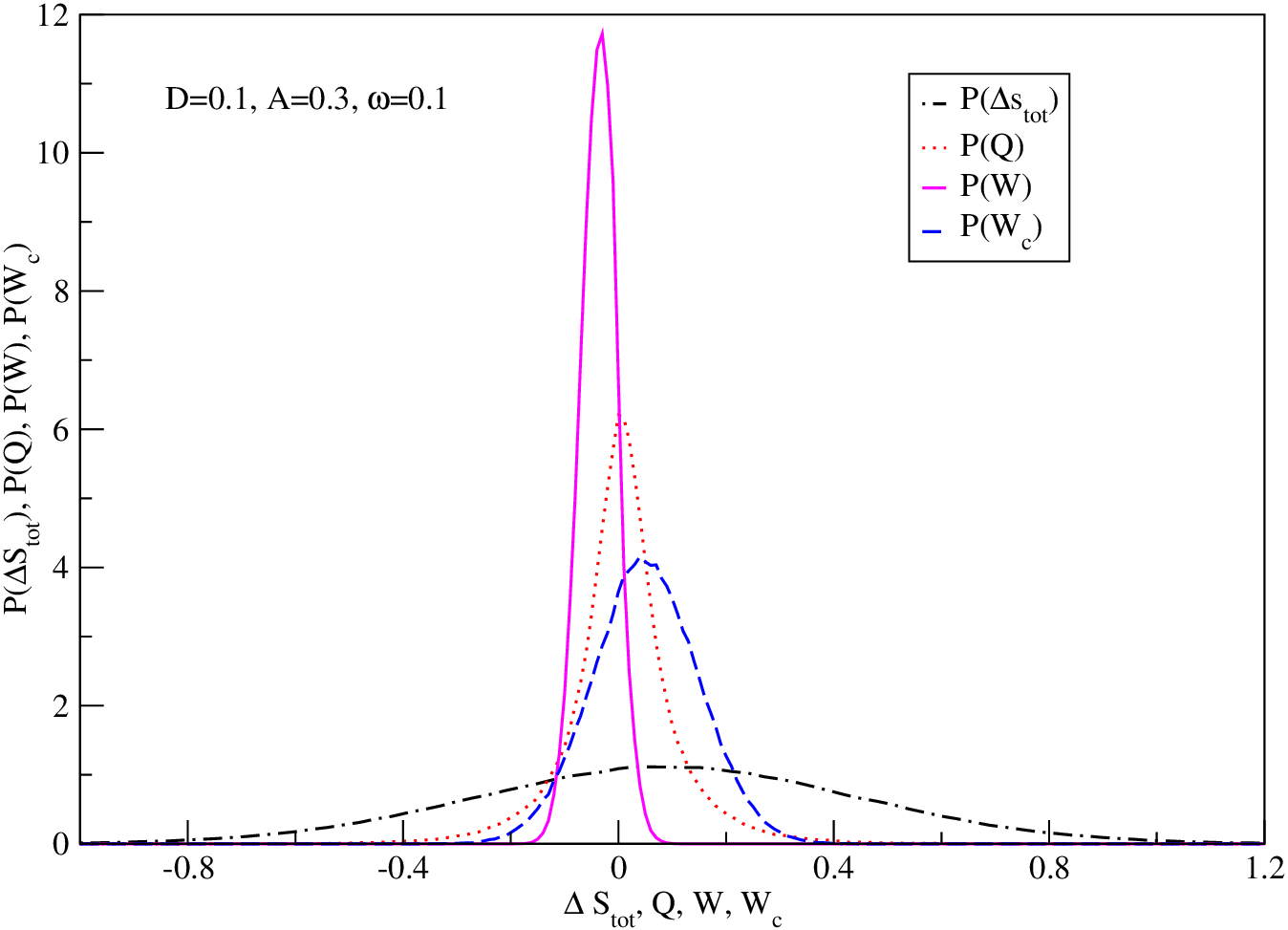}
\caption{The distributions $P(\Delta S_{tot})$, $P(W)$, $P(W_{c})$ and $P(Q)$ for the parameters $A=0.3$, $D=0.1$, $\omega=0.1$ and for $t=\frac{T_{p}}{4}$.} 
\end{center}
\end{figure}
To put the probability distributions of physical quantities $W$, $W_{c}$, $Q$ and $\Delta S_{tot}$ in the same perspective we have generated $10^{5}$ trajectories of Brownian particle. Total entropy production $\Delta S_{tot}$ comprises of two parts namely system entropy and medium entropy. For details we refer to [18,19]. For each stochastic trajectory we calculate $W$, $W_{c}$, $Q$ and $\Delta S_{tot}$ and using these values, the obtained $P(W)$, $P(W_{c})$, $P(Q)$ and $P(\Delta S_{tot})$ are plotted in fig (4). Areas under the curves for $W<\Delta F$, $W_{c}<0$, $Q<0$ and $\Delta S_{tot} <0$ are different. In our linear problem the distribution for $P(W)$, $P(W_{c})$ and $P(\Delta S_{tot})$ are Gaussian, while that for $P(Q)$ is non Gaussian. It may be noted that by appropriately choosing the system and protocol  most probable value of $P(\Delta S_{tot})$ can be shifted to the negative side, yet $\langle \Delta S_{tot} \rangle >0$. However, peak in $P(Q)$ corresponding to the most probable value occurs at positive values of $Q$ only. Similar behaviour in regard to $P(W)$ and $P(W_{c})$ are pointed earlier. In the next section, we study the nature of fluctuations in $W$ and $W_{c}$ for a nonlinear driven system in the time periodic steady state. 
\begin{figure}[htp!]
\begin{center}
\input{epsf}
\includegraphics [width=4.5in,height=3.5in] {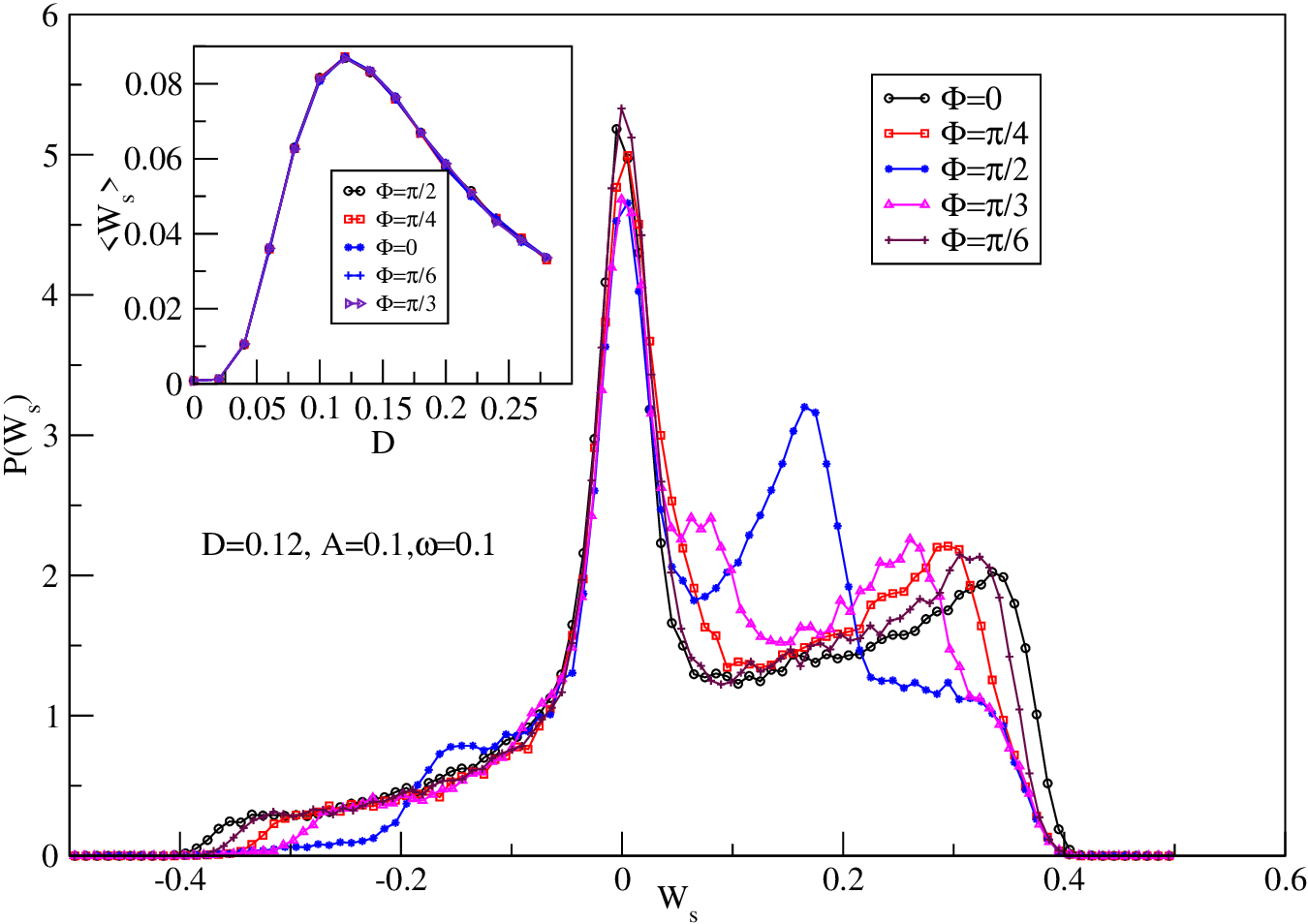} 
\caption{The distribution $P(W_{s})$ over a single period for various values of initial phase $\phi$. The other parameters are $A=0.1$, $D=0.12$ and $\omega=0.1$. Inset shows plot of $\langle W_{s} \rangle$ vs $D$ for different value of $\phi$ for the same physical parameters.} 
\end{center}
\end{figure}
\section{Driven nonlinear systems}
To this end we consider the motion of a Brownian particle in a bistable system under the action of an external ac force.~~~The total  potential is given by $U(x,t)=U_{0}(x)+U_{p}(x,t)$.~~The static double well potential is $U_{0}(x)=-\frac{1}{2}x^{2}+\frac{1}{4}x^{4}$ and the time dependent potential is $U_{p}(x,t)=-Ax(t)\cos(\omega t+\phi)$.~~This system is shown to exhibit the well known phenomenon of stochastic resonance (SR) [22-28]. Theoretical study on thermodynamic work and heat have been shown to satisfy SSFT [26,27]. Experiments in connection with fluctuation theorems have been carried out in systems exhibiting SR [28,29].~~We have numerically evaluated the classical work and thermodynamic work distributions over a single period/large number of periods.~~For description of numerical method we refer to [26,27].
\begin{figure}[htp!]
\begin{center}
\input{epsf}
\includegraphics [width=4.5in,height=3.5in] {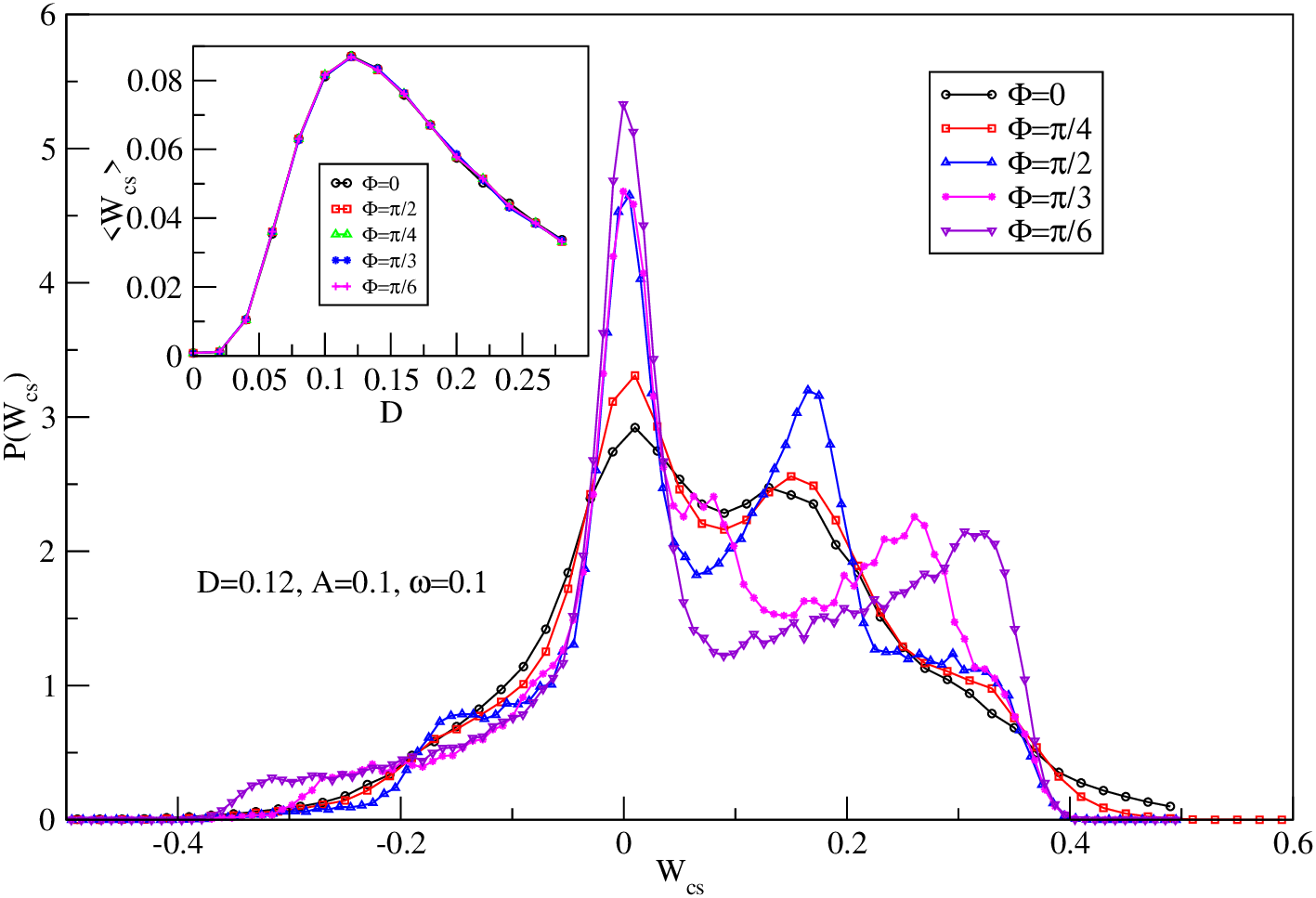} 
\caption{The distribution $P(W_{cs})$ over a single period for various values of initial phase $\phi$. The other parameters are $A=0.1$, $D=0.12$ and $\omega=0.1$. Inset shows plot of $\langle W_{cs} \rangle$ vs $D$ for different values of $\phi$ for the same physical parameters.} 
\end{center}
\end{figure}
\begin{figure}[htp!]
\begin{center}
\input{epsf}
\includegraphics [width=4.5in,height=3.5in] {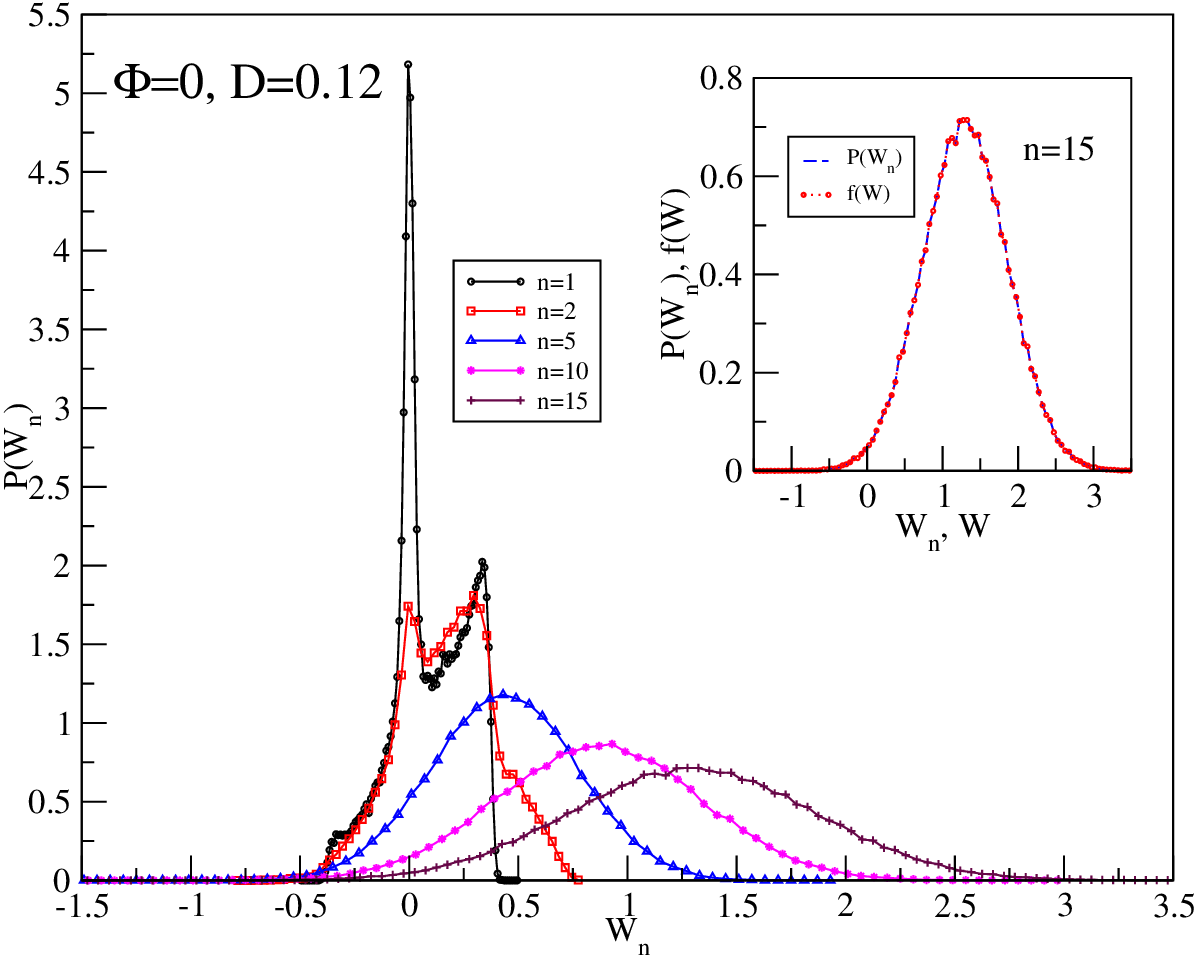} 
\caption{The distributions $P(W_{n})$ for different periods ($n$) in the time periodic steady state. In the inset $P(W_{n})$ is plotted along with Gaussian fit $f(W)$ for $n=15$. From the Gaussian fit $f(W)$, the fluctuation ratio has been calculated as $\frac{\sigma^{2}}{\frac{2}{\beta}\langle W_{n} \rangle}=1.022$. Here $D=0.12$, $A=0.1$ and $\omega=0.1$.} 
\end{center}
\end{figure}

In fig(5) and (6) we have plotted probability distributions of thermodynamic work $W_{s}$ and classical work $W_{cs}$ evaluated over a single period in the time periodic regime for various values of initial phases. For a given fixed phase, distributions $P(W_{s})$ and $P(W_{cs})$ are different. Only for phase $\phi=\frac{\pi}{2}$, both distributions are identical. For this particular phase, it can be readily seen that thermodynamic work and classical work are identical when evaluated over a period(s). The multipeaked structure  in the  distributions arise from the interwell and intrawell dynamics of the particle in a double well system. For details see Ref.[27,28]. Unlike the sensitivity of $P(W_{cs})$ and $P(W_{s})$, $\langle W_{s} \rangle$ and $\langle W_{cs} \rangle$ over a cycle do not depend on the  initial phase and are identical, i.e., $\langle W_{cs} \rangle=\langle W_{s} \rangle$. They are plotted in the inset of fig(5) and (6). $\langle W_{s} \rangle$($=\langle W_{cs} \rangle$) exhibits a peak as a function of noise strength $D$. This phenomenon is referred to as SR [25,26,27]. At the value of noise strength corresponding to the peak value in $\langle W_{c} \rangle$, the random hops of the Brownian particle between the two wells get synchronized with the external drive[22,28]. 

\begin{figure}[htp!]
\begin{center}
\input{epsf}
\includegraphics [width=4.5in,height=3.5in] {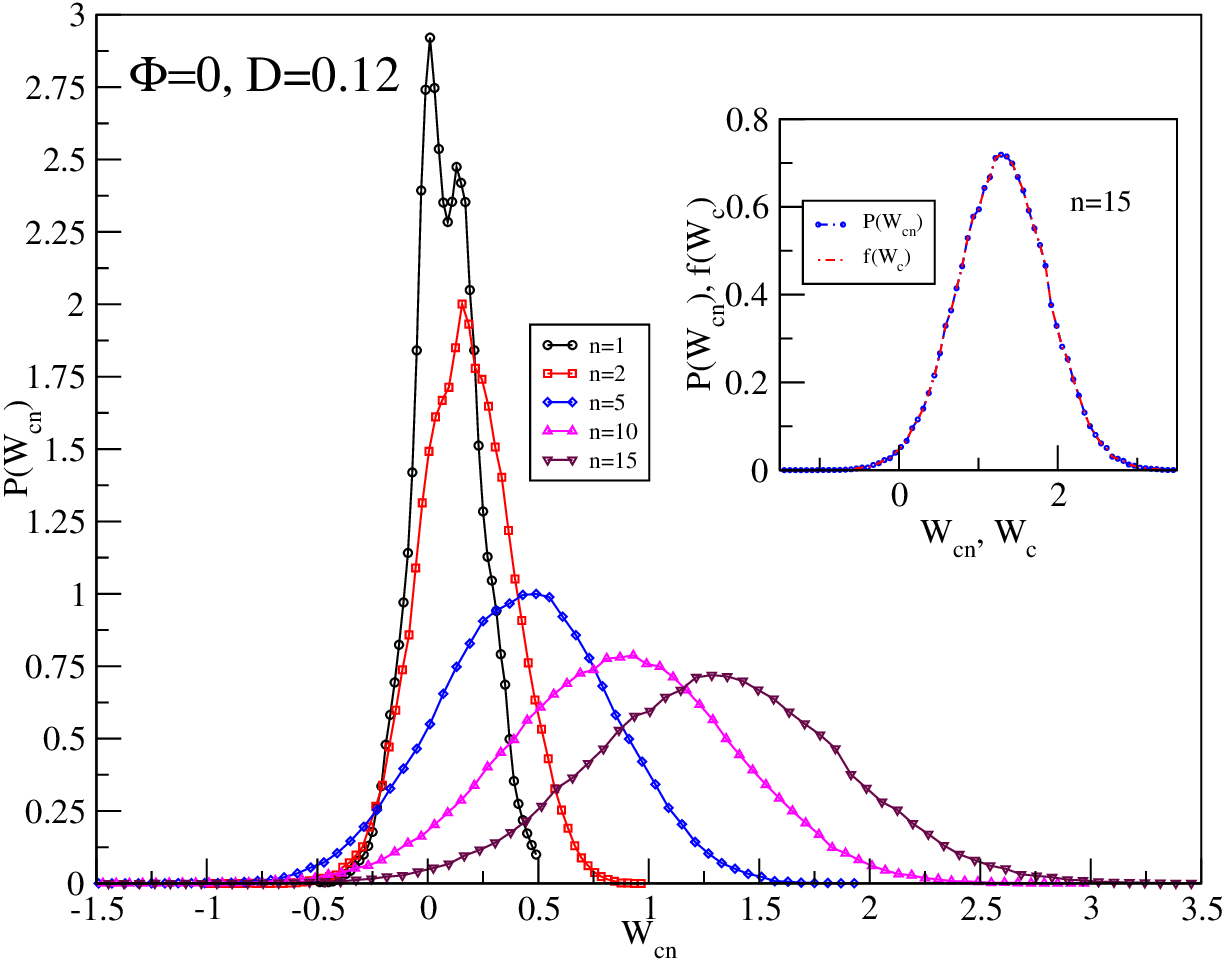} 
\caption{The distributions $P(W_{cn})$ for $n$ periods versus $W_{cn}$ in the time periodic steady state. Inset shows the plot of $P(W_{cn})$ along with it's Gaussian fit $f(W_{c})$ for $n=15$. From the Gaussian fit $f(W_{c})$, the fluctuation ratio has been calculated as $\frac{\sigma^{2}}{\frac{2}{\beta}\langle W_{cn} \rangle}=1.005$. The other parameters are same as in fig(7).} 
\end{center}
\end{figure} 

In fig(7) and (8), we have plotted $P(W_{n})$ and $P(W_{cn})$ versus thermodynamic work done over $n$ cycles, $W_{n}$ and classical work done over $n$ cycles, $W_{cn}$ respectively. In both these cases, with increasing $n$, multipeaked distributions become smoother and for large $n$, they both evolve towards Gaussian distribution. Corresponding Gaussian fit for $P(W_{n})$ and $P(W_{cn})$ for $n=15$ are shown in the inset of fig(7) and (8) respectively. For $P(W_{n})$ and for $n=15$, from the Gaussian fit we get variance $\sigma^{2}=0.3206$, $\langle W_{n} \rangle=1.30701$ and the corresponding fluctuation ratio, $\frac{\sigma^{2}}{\frac{2}{\beta}\langle W_{n} \rangle}=1.022$, which is close to unity (within our numerical accuracy). The Gaussian nature of $W_{cn}$ along with fluctuation ratio being $1$ implies validity of SSFT. Similar conclusions can be made for $P(W_{cn})$ for $n=15$, where the Gaussian fit gives $\sigma^{2}=0.31525$, and hence the fluctuation ratio (1.005) is close to unity. The presence of non-Gaussian tails at large value of $W_{cn}$ and $W_{n}$ are not ruled out. However, numerically it is difficult to detect them. It may also be noted that time interval after which SSFT is obeyed for $W$ and $W_{c}$ may be different and depends of the physical parameters.
\section{Conclusion}
We have studied fluctuations in $W$ and $W_{c}$ analytically for both transient and time periodic steady states in case of a linear model. We have also shown that as the observation time of trajectories increases, the number of trajectories which exhibit atypical behaviour (namely $W_{c}$ and $Q$ being negative, $W$ being less than $\Delta F$) decreases. Thus in the limit of large time of observations, macroscopic thermodynamic behaviour results for physical quantities. We have discussed the validity of SSFT for both $W$ and $W_{c}$ in our linear as well as nonlinear models. 

\section{Acknowledgment}
One of us (AMJ) thanks DST,~~India for financial support. MS thanks Mr. Sourabh Lahiri for his help in computation and Mr. Devashish Sanyal for useful discussions.

\section{Appendix-A}
\begin{eqnarray}
\langle W \rangle &=&\frac{A^2 k}{4(k^2+\omega^2)}[\cos(2(\omega t+\phi))-\cos(2\phi)] \nonumber\\
&+& \frac{A^2\omega^2 t}{2(k^2+\omega^2)}+\frac{A^2\omega}{4(k^2+\omega^2)}[\sin(2(\omega t+\phi)-\sin(2\phi)] \nonumber\\
&+& \frac{A^{2}\omega}{(k^{2}+\omega^{2})^{2}} \exp(-kt)[-k\cos(\omega t+\phi)+\omega \sin(\omega t+\phi)](k\sin(\phi)-\omega\cos(\phi))\nonumber\\&+&\frac{A^{2}\omega}{(k^{2}+\omega^{2})^{2}} (k\sin(\phi)-\omega \cos(\phi))(k\cos(\phi)-\omega \sin(\phi))\nonumber\\&-&\frac{A^{2} \omega \sin(\phi)}{k^{2}+\omega^{2}}[\exp(-kt)\{-k\cos(\omega t+\phi)\nonumber\\&+&\omega\sin(\omega t+\phi)\}+(k\cos(\phi)-\omega \sin(\phi))]
\end{eqnarray}
\section{Appendix-B}
\begin{eqnarray}
\langle W_{c} \rangle &=&\frac{A^2 k}{4(k^2+\omega^2)}[\cos(2(\omega t+\phi))-\cos(2\phi)] \nonumber\\
&+& \frac{A^2\omega^2 t}{2(k^2+\omega^2)}+\frac{A^2\omega}{4(k^2+\omega^2)}[\sin(2(\omega t+\phi)-\sin(2\phi)] \nonumber\\
&+& \frac{A^{2}\omega}{(k^{2}+\omega^{2})^{2}} \exp(-kt)[-k\cos(\omega t+\phi)+\omega \sin(\omega t+\phi)](k\sin(\phi)-\omega\cos(\phi))\nonumber\\&+&\frac{A^{2}\omega}{(k^{2}+\omega^{2})^{2}} (k\sin(\phi)-\omega \cos(\phi))(k\cos(\phi)-\omega \sin(\phi))\nonumber\\&+&\frac{A^{2}k}{k^{2}+\omega^{2}}\sin^{2}(\omega t+\phi)-\frac{A^{2}\omega}{2(k^{2}+\omega^{2})}\sin(2(\omega t+\phi))\nonumber\\&+&\frac{A^{2}}{k^{2}+\omega^{2}}\exp(-kt)\sin(\omega t+\phi)[\omega \cos(\phi)-k\sin(\phi)].
\end{eqnarray}
\section{Appendix-C}
The expression for $\Delta(\tau)$ is given by
\begin{eqnarray}
\Delta(\tau)=\frac{2}{\beta}\langle W_{\tau}\rangle-\frac{2A^{2}{\omega^{2}}}{k\beta(k^{2}+\omega^{2})^{2}}(k^{2}\cos^{2}(\phi)-\omega^{2}\sin^{2}(\phi))(1-\exp(-k\tau))
\end{eqnarray}
\section{Appendix-D}
The analytical expression for $\Delta_{1}(\tau)$ is given by
\begin{eqnarray}
\Delta_{1}(\tau)&=&\frac{2}{\beta}\langle W_{c\tau} \rangle +\frac{2A^{2}\omega^{2}}{k\beta(k^{2}+\omega^{2})^{2}}(k^{2}\cos^{2}(\phi)-\omega^{2}\sin^{2}(\phi))(1+\exp(-k\tau))\nonumber\\&-&\frac{4A^{2}\omega^{2}k}{\beta (K^{2}+\omega^{2})^{2}}+\frac{2A^{2}}{k\beta}\sin^{2}(\phi)+\frac{2A^{2}}{k\beta}\frac{(k^{2}-\omega^{2})}{(k^{2}+\omega^{2})}\sin^{2}(\phi)\exp(-k\tau)
\end{eqnarray}

\end{document}